\documentclass[prb,twocolumn,superscriptaddress,showpacs,amsmath,amssymb,floatfix]{revtex4-1}

\usepackage{graphicx}
\usepackage{latexsym}
\usepackage{amsmath}
\usepackage{amssymb}
\usepackage{amsfonts}
\usepackage{color}
\usepackage{bm}
\usepackage{verbatim}
\usepackage[]{color}
\usepackage{float}
\definecolor{red}{rgb}{1,0,0}
\usepackage{framed}
\definecolor{shadecolor}{RGB}{222,222,221}
\graphicspath{{Figs'/}}

\definecolor{MS-color}{RGB}{128,0,128}

\begin{document}

\title{Electrical response of S-F-TI-S junctions on  magnetic texture dynamics}

 \date{\today}
 
\author{D. S. Rabinovich}
\affiliation{Moscow Institute of Physics and Technology, Dolgoprudny, 141700 Russia}
\affiliation{Skolkovo Institute of Science and Technology, Skolkovo 143026, Russia}
\affiliation{Institute of Solid State Physics, Chernogolovka, Moscow
  reg., 142432 Russia}

\author{I. V. Bobkova}
\affiliation{Institute of Solid State Physics, Chernogolovka, Moscow
  reg., 142432 Russia}
\affiliation{Moscow Institute of Physics and Technology, Dolgoprudny, 141700 Russia}
\affiliation{National Research University Higher School of Economics, Moscow, 101000 Russia}

\author{A. M. Bobkov}
\affiliation{Institute of Solid State Physics, Chernogolovka, Moscow reg., 142432 Russia}

\begin{abstract}
We consider a hybrid structure consisting of superconducting or normal leads with a combined ferromagnet-3D topological insulator interlayer. We compare responses of a Josephson junction and a normal junction to magnetic texture dynamics. In both cases the electromotive force resulting from the magnetization dynamics generates a voltage between the junction leads. For an open circuit this voltage is the same for normal and superconducting leads and allows for electrical detection of magnetization dynamics and a structure of a given magnetic texture. However, under the applied current the electrical response of the Josephson junction is essentially  different due to the strong dependence of the critical Josephson current on the magnetization direction  and can be used for experimental probing of this dependence. We propose a setup, which is able to detect a defect motion and to provide detailed information about the structure of magnetic inhomogeneity. The discussed effect could be of interest for spintronics applications. 
\end{abstract}

  \maketitle
 
\section{Introduction}
 
At present the magnetic field-driven and current-driven switching in spin valves and related systems, as well as domain-wall (DW) and skyrmion motion is the focus of research activity, which
is in part motivated by the promising application potential for
spintronics.  The majority of the corresponding literature is devoted to the dynamics generation: structure of the driving torques and their ability to induce the dynamics. On the contrary, the purpose of present work is to study prospects of electrical detection of a {\it given} magnetization dynamics, which can be caused either by an applied magnetic field or by application of the electric current or by a magnonic torque. The key element of the detection scheme is a hybrid structure consisting of probing leads and the ferromagnet in a combination with a 3D topological insulator (3D TI), where the spin-momentum locking of 3D topological insulator surface states provides a conversion of the magnetization dynamics in the ferromagnet into an electric voltage.  We demonstrate that the suggested method of  electrical detection is able to provide detailed information about the structure of time-dependent magnetization texture. The physical basis of the effect is the so-called electromotive force (emf) induced by the magnetization dynamics. The emf has been widely discussed in literature in the context of magnetization dynamics in metallic ferromagnets \cite{Stern1992,Stone1996,Volovik1987,Berger1986,Barnes2007,Duine2008,Saslow2007,Tserkovnyak2009,Zhang2009,Yang2009,Yang2010}. Due to the existence of the electromotive force the DW motion leads to appearance of an additional voltage drop in the region occupied by the moving wall. This voltage can vary in the range from $nV$ to $\mu V$ \cite{Yang2009} and in special situations can be used for electrical detection of the presence of magnetization dynamics \cite{Barnes2006}. The emf can be considered as a consequence of the presence of a time-dependent gauge potential in the local spin basis of a spin-textured system\cite{Berry1984,Volovik1987}.

The spin-orbit coupling can also be described in terms of SU(2) gauge potential\cite{Frohlich1993,Rebei2006,Jin2006,Jin2006_2,Bernevig2006,Hatano2007,Leurs2009,Tokatly2008,Bergeret2013}, which becomes time-dependent in the local spin basis in the presence of a magnetization dynamics and, therefore, also results in the appearance of the emf \cite{Kim2012,Tatara2013,Yamane2013,Rabinovich2019}. 
The property of spin-momentum locking\cite{Burkov2010,Culcer2010,Yazev2010,Li2014} of surface states of a 3D TI can be viewed as an extremely strong spin-orbit coupling. Proximity to a ferromagnet induces an effective exchange field in the 3D TI surface states, which follows the dynamics of the ferromagnet magnetization. The combination of the spin-momentum locking and the induced exchange field results in the appearance of a special type of the emf in the 3D TI surface layer, which is determined by time derivatives of the in-plane magnetization components \cite{Garate2010,Nomura2010,Tserkovnyak2012}. Motivated by this fact here we consider a hybrid structure in a Josephson junction geometry L/(F/TI)/L, where L is a lead electrode, which can be as normal (N), so as superconducting (S), and F/TI is an interlayer consisting of a ferromagnet (F) and a 3D topological insulator (TI). Here we investigate the electrical voltage generated by externally induced magnetization dynamics at the junction and prospects of the effect for electrical probing of time-dependent magnetization structure. We compare the responses of a Josephson junction and a normal junction to the magnetization dynamics. In both cases the electromotive force resulting from the magnetization dynamics generates a voltage between the junction leads. In the open circuit geometry the voltage is the same for both normal and superconducting leads and allows for electrical detection of magnetization dynamics. However, under the applied current the electrical response of the Josephson junction is essentially  different due to the strong dependence of the critical Josephson current on the magnetization direction and can be used for experimental probing of this dependence.

It is timely to study F/3D TI hybrids because at present there is great progress in their experimental realization. In particular, to introduce the ferromagnetic order into the TI, random doping of transition metal elements, e.g., Cr or V, has been employed \cite{Chang2013,Kou2013,Kou2013_2,Chang2015}. The second option, which has been successfully realized experimentally, is a coupling of a nonmagnetic TI to a high $T_c$
magnetic insulator to induce strong exchange interaction in the surface states via the proximity effect\cite{Jiang2015,Wei2013,Jiang2015_2,Jiang2016}. The spin injection into a TI surface states via the spin pumping techniques has also been realized and the resulting spin-electricity conversion has been measured \cite{Shiomi2014,Deorani2014,Jamali2015,Rojas-Sanchez2016}.

The investigated electrical response of the Josephson junction to the magnetization dynamics could be of interest for spintronics applications because it provides a way to electrically read information encoded  in the magnetization. We propose a setup, which is able not only to detect a defect motion but also to provide detailed information about the structure of magnetic inhomogeneity.

The paper is organized as follows. In Sec.II we describe the system under consideration, formulate the necessary equations and calculate the general expression for the electric current in  the presence of magnetization dynamics. In Sec.III we investigate the electrical response of the system to magnetization dynamics in two particular situations: for an open circuit and in the presence of the constant applied electric current. In that section we also demonstrate the possibilities of using the effect to detect DW motion and DW structure. Our conclusions are formulated in Sec.IV.

\section{Model and method} 

\begin{figure}[htb!]
 \centerline{\includegraphics[clip=true,width=2.6in]{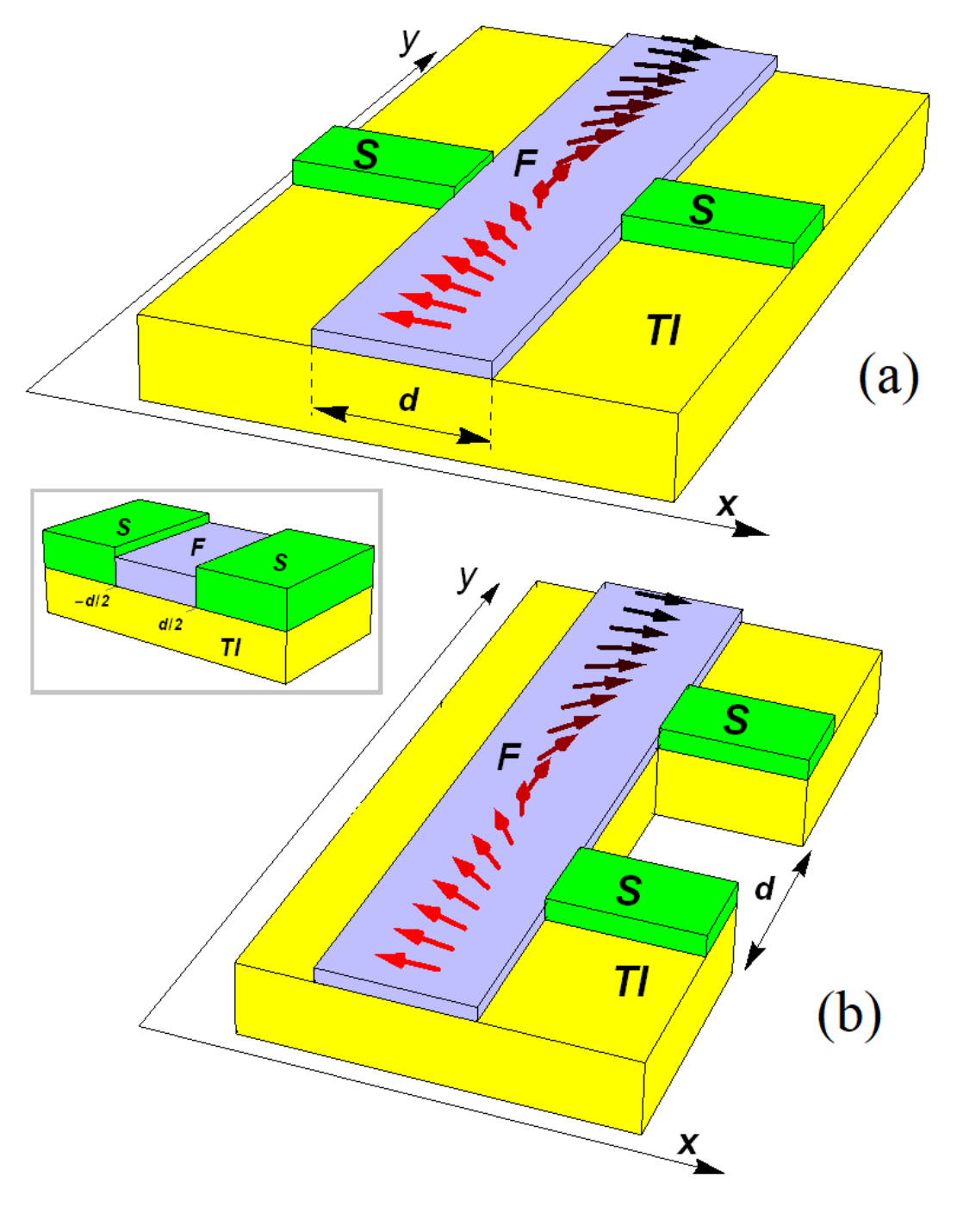}}
  \caption{(a) and (b) are possible sketches of the system under consideration. Insert: detector regions in the both panels are modelled by a S/F-TI/S junction.}
       \label{sketch}
 \end{figure}

The possible sketches of the system under consideration are presented in Fig.~\ref{sketch}. The  interlayer region of a S/3D TI/S Josephson (or N/3D TI/N) detecting  junction is covered by a ferromagnet. The magnetization dynamics is assumed to be induced in the ferromagnet by external means. We discuss  particular examples below, and now we derive a formalism allowing for studying the electrical response to the magnetization dynamics of a general type and consider only the detector region, which can be viewed as a S(N)/3D TI/S(N) junction (see insert to Fig.~\ref{sketch}, for concreteness $x$-axis is chosen perpendicular to S/F interfaces). In principle, the ferromagnet can be as metallic, so as insulating, but insulating ferromagnets are better candidates for the measurements, as it is explained below. We assume that the transport between the leads occurs via the TI surface states. This is strictly the case for insulating ferromagnets. We believe that our results can be of potential interest for systems based on $Be_2Se_3/YIG$ or $Be_2Se_3/EuS$ hybrids, which were realized experimentally \cite{Jiang2015,Wei2013,Jiang2015_2,Jiang2016}. For metallic ferromagnets the situation is a bit more complicated. If the ferromagnet is strong, that is its exchange field is comparable to the Fermi energy, then the Josephson current through the ferromagnet is greatly suppressed and indeed flows through the 3D TI surface states. At the same time the normal (quasiparticle) current mainly flows in the ferromagnet because its resistance is typically much smaller than the resistance of the conducting surface layer of the TI. This provides an additional channel for the normal current, which prevents the observation of the effect under consideration, as discussed below. 

It is assumed that the magnetization $\bm M(\bm r)$ of the ferromagnet induces an effective exchange field $\bm h_{eff}(\bm r) \sim -\bm M (\bm r)$ in the underlying conductive TI surface layer and $h_{eff}$ is small as compared to the exchange field in the ferromagnet. The Hamiltonian that describes the TI surface states in the
presence of an in-plane exchange field $\bm h_{eff}(\bm r)$ reads:
\begin{equation}
\hat H=\int d^2 \bm r \hat \Psi^\dagger (\bm r)\hat H(\bm r)\hat \Psi(\bm r)
\label{H},
\end{equation}
\begin{equation}
\hat H(\bm r)=-iv_F (\bm \nabla \times \bm e_z)\hat {\bm \sigma}+\bm h_{eff}(\bm r)\hat {\bm \sigma} -\mu
\label{h},
\end{equation}
where $\hat \Psi=(\Psi_\uparrow, \Psi_\downarrow)^T$, $v_F$ is
the Fermi velocity, $\bm e_z$ is a unit vector normal to the surface
of the TI, $\mu$ is the chemical potential, and $\hat {\bm \sigma}=(\sigma_x, \sigma_y, \sigma_z)$ is a vector of
Pauli matrices in spin space. It was shown \cite{Schwab2011,Zyuzin2016,Bobkova2016} that in the quasiclassical approximation $(h_{eff},\varepsilon) \ll \mu$ the Green's function has the following spin structure: 
\begin{equation}
\check g (\bm n_F, \bm r, \varepsilon,t) = \hat g (\bm n_F, \bm r, \varepsilon,t) (1+\bm n_\perp \bm \sigma)/2,
\label{spin_structure}
\end{equation} 
where $\bm n_\perp = (n_{F,y},-n_{F,x},0)$ is the unit vector perpendicular to the direction of the quasiparticle trajectory $\bm n_{F} = \bm p_F/p_F$ and $\hat g$ is the {\it spinless} $4 \times 4$ matrix in the direct product of particle-hole and Keldysh spaces. The spin structure above reflects the fact that the spin and momentum of a quasiparticle at the surface of a 3D TI are strictly locked and make a right angle.

 Here we assume that the TI surface states are in the ballistic limit because this regime is more relevant for existing experiments. Following standard procedures\cite{Eilenberger1968,Usadel1970} it was 
demonstrated\cite{Schwab2011,Zyuzin2016,Bobkova2016,Hugdal2017} that the  
spinless Green's function $\hat g(\bm n_F, \bm r, \varepsilon,t)$ obeys the 
following transport equations in the ballistic limit:
\begin{eqnarray}
-i v_F \bm n_F \hat \nabla \hat g = \Bigl[ \varepsilon \tau_z - \hat \Delta, \hat g \Bigr]_\otimes,
\label{eilenberger}
\end{eqnarray}
where the spin-momentum locking allows for including $\bm h_{eff}$ into the gauge covariant gradient
\begin{equation}
\hat{\nabla} \hat g = \bm \nabla \hat g + (i/v_F)[(h_x \bm e_y - h_y \bm e_x), \hat g]_\otimes.
\label{covariant_gradient}
\end{equation}
$[A,B]_\otimes = A\otimes B -B \otimes A$ and $A \otimes B = \exp[(i/2)(\partial_{\varepsilon_1} \partial_{t_2} -\partial_{\varepsilon_2} \partial_{t_1} )]A(\varepsilon_1,t_1)B(\varepsilon_2,t_2)|_{\varepsilon_1=\varepsilon_2=\varepsilon;t_1=t_2=t}$. $\tau_{x,y,z}$ are Pauli matrices in particle-hole space with $\tau_\pm = (\tau_x \pm i \tau_y)/2$. $\hat \Delta = \Delta(x)\tau_+ - \Delta^*(x)\tau_-$ is the matrix structure of the superconducting order parameter $\Delta(x)$ in the particle-hole space. We assume $\Delta(x)=\Delta e^{-i\chi/2}\Theta(-x-d/2)+\Delta e^{i\chi/2}\Theta(x-d/2)$.

Eq.~(\ref{eilenberger}) should be supplemented by the normalization condition $\hat g\otimes \hat g = 1$ and  boundary conditions at $x=\mp d/2$. As a minimal model we assume no potential barriers at the $x=\mp d/2$  interfaces and consider these interfaces as fully transparent.  In this case the boundary conditions are extremely simple and are reduced to continuity of $\hat g$ for a given quasiparticle trajectory at the interfaces. However, our result for the emf remains valid even in a diffusive case and for low-transparent interfaces. 

The density of electric current along the $x$-axis can be calculated via the Green's function as follows:
\begin{eqnarray}
j_x = -\frac{e N_F v_F}{4} \int \limits_{-\infty}^{\infty} d \varepsilon \int \limits_{-\pi}^{\pi} \frac{d \phi}{2 \pi} \cos \phi g^K,
\label{current}
\end{eqnarray}
where $\phi$ is the angle the quasiparticle trajectory makes with the $x$-axis. $g^K$ is the Keldysh part of the normal Green's function, which can be expressed via the retarded, advanced parts and the distribution function $\varphi$ as follows: $g^K = g^R \otimes \varphi - \varphi \otimes g^A$. 
In general, the electric current through the junction consists of two parts: the Josephson current $j_s$ and the normal current $j_n$. The Josephson current is connected to the presence of the nonzero anomalous Green's functions in the interlayer and exists even in equilibrium. Below we calculate the both contributions to the current microscopically. It is assumed that the effective exchange field in the interlayer of the junction is spatially homogeneous. 

\begin{figure}[htb!]
\begin{minipage}[b]{\linewidth}
 \centerline{\includegraphics[clip=true,width=3.0in]{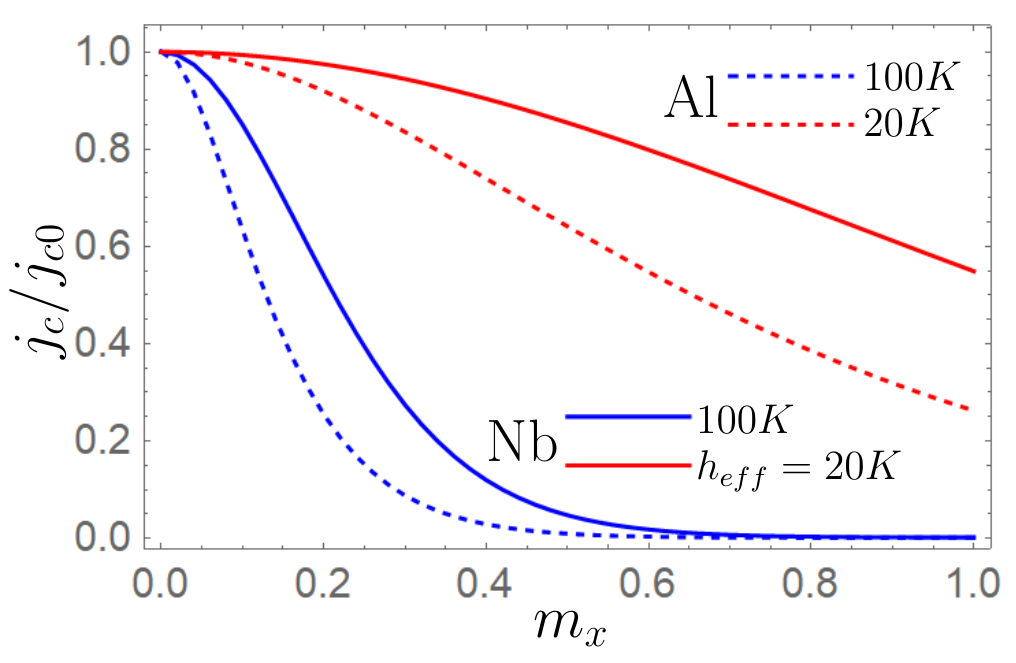}}
 \end{minipage}\hfill
 \caption{$j_c$ as a function of $m_x$ for $r=13.2$, $d/\xi_N =4.1 $ (solid blue); $r=2.6$, $d/\xi_N = 4.1$ (solid red);  $r=13.2$, $d/\xi_N =0.74 $ (dashed blue); $r=2.6$, $d/\xi_N =0.74 $ (dashed red). $j_c$ is normalized to $j_{c0} \equiv j_c(m_x=0)$.}
       \label{current_orientation}
 \end{figure}

Here we work near the critical temperature of the superconductors, that is $\Delta/T_c \ll 1$. The Josephson current for the system under consideration in this regime has already been calculated \cite{Nashaat2019} and takes the form:
\begin{eqnarray}
j_s = j_c \sin (\chi - \chi_0), \label{Josephson_CPR}\\
j_c = j_b \int \limits_{-\pi/2}^{\pi/2} d \phi \cos \phi  \times \nonumber \\
\exp\Bigl[-\frac{2\pi T d}{v_F \cos \phi}\Bigr] \cos \Bigl[\frac{2h_xd \tan \phi}{v_F}\Bigr], 
\label{critical_current_T} \\
\chi_0 = 2 h_y d/v_F \label{chi_0},
\end{eqnarray}
where $j_b = ev_F N_F \Delta^2/(\pi^2 T)$. Similar expression has already been obtained for Dirac materials \cite{Hugdal2017}. It is seen from Eqs.~(\ref{critical_current_T})-(\ref{chi_0}) that the Josephson current manifest strong dependence on the orientation of the ferromagnet magnetization. It is sensitive to the $y$-component of the magnetization only via the anomalous phase shift \cite{Tanaka2009,Linder2010,Zyuzin2016}, therefore this component does not lead  to superconductivity suppression in the interlayer and does not influence the amplitude of the critical current. At the same time $x$-component of the magnetization does not couple to superconductivity via the anomalous phase shift, but causes the superconductivity depairing in the interlayer leading to the suppression of the critical current. The suppression of the critical current as a function of $m_x \equiv M_x/M_s$ is presented in Fig.~\ref{current_orientation}. For estimates we take  $d=50nm$, $v_F = 10^5m/s$  and $T_c = 10K$, what corresponds to the parameters of $Nb/Bi_2Te_3/Nb$ Josephson junctions\cite{Veldhorst2012}. In this case $\xi_N = v_F/2\pi T_c \approx 12nm$. We have also plotted $j_c(m_x)$ for $T_c = 1.8K$, what corresponds to the Josephson junctions with $Al$ leads. 

It is difficult to give an accurate a-priori estimate of $h_{eff}$ because there are no reliable experimental data on its value. However, basing on the experimental data on the Curie temperature of the magnetized TI surface states \cite{Jiang2015_2}, where the Curie temperature in the range $20-150K$ was reported, we can roughly estimate $h_{eff} \sim 0.01- 0.1 h_{YIG}$. We assume $h_{eff} \sim 20-100 K$ in our numerical simulations, what corresponds to the dimensionless parameter $r = 2 h_{eff} d/v_F = 2.6-13.2$. The both parts of the strong dependence of the Josephson current on the magnetization orientation (the dependence via anomalous phase shift and the dependence via the amplitude of the critical current) manifest themselves in the electrical response of the junction on the magnetization dynamics, as we demonstrate below.

The normal current is due to the deviation of the distribution function from the equilibrium. It was not calculated in Ref.~\onlinecite{Nashaat2019}, therefore we discuss it in detail. In order to find $\varphi$ we have to solve the kinetic equation, which can be obtained from the Keldysh part of Eq.~(\ref{eilenberger}) and takes the form
\begin{eqnarray}
-v_{F,x} \partial_x \varphi = \dot \varphi + \dot {\bm h}_{eff} \bm n_\perp \partial_\varepsilon \varphi. 
\label{kinetic_equation} 
\end{eqnarray}
When deriving this equation we have made the following assumptions: 1) we have neglected all the corrections to the kinetic equation due to superconductivity, because they lead to the corrections in the final expression for the normal current of the order of $(\Delta/T_c)^2$, which can be safely neglected near the critical temperature. In this approximation we neglect the terms of the same order $(\Delta/T_c)^2(eV/T_c)$ in the Josephson current, as well as in the normal current; 2) the interlayer of the junction is assumed to be shorter than the inelastic energy relaxation length, therefore all the inelastic relaxation processes are neglected in Eq.~(\ref{kinetic_equation}) and 3) we have also expanded $\otimes$-products up to the lowest order with respect to time derivatives: $A \otimes B \approx AB +(i/2)(\partial_\varepsilon A \partial_t B - \partial_t A \partial_\varepsilon B)$. The applicability of this expansion is justified by the fact that the voltage, induced by the magnetization dynamics at the junction is small $eV/(k_B T_c) \ll 1$, what is demonstrated by further numerical calculations.

We solve Eq.~(\ref{kinetic_equation}) neglecting the term $\dot \varphi$ and assuming that the deviation of the distribution function from equilibrium is small $\varphi = \tanh (\varepsilon/2T)+ \delta \varphi$. The last assumption is justified by the condition $eV/k_BT_c \ll 1$. The term $\dot \varphi$ can be neglected if $d/(v_F t_d) \ll 1$, where $t_d$ is the characteristic time of magnetization variations. For estimates we take $d=50nm$, $v_F = 10^5m/s$ and $t_d = 0.5 \times 10^{-8}s$ (this value of $t_d$ corresponds to material parameters of YIG thin films, see below). In this case $d/(v_F t_d) \sim 10^{-4}$ and the term $\dot \varphi$ can safely be neglected.

The solution should also satisfy the asymptotic values $\varphi_{\pm}= \tanh[(\varepsilon \mp eV/2)/2T] \approx \tanh(\varepsilon/2T) \mp (eV/4T)\cosh^{-2}(\varepsilon/2T)$ at $x = \mp d/2$ if we assume that the leads are in equilibrium except for the voltage drop $V$ between them. In this case the solution of Eq.~(\ref{kinetic_equation}) can be easily found and takes the form:
\begin{eqnarray}
\varphi_\pm =  \tanh\frac{\varepsilon}{2T} \mp \frac{eV}{4T}\frac{1}{\cosh^{2}(\varepsilon/2T)} - \nonumber \\ \frac{(x \pm d/2)}{2T \cosh^{2}(\varepsilon/2T)}\frac{\dot {\bm h}_{eff} \bm n_\perp }{v_{F,x}},
\label{kinetic_sol} 
\end{eqnarray}
where the subscript $\pm$ corresponds to the trajectories ${\rm sgn}\! ~v_x =\pm 1$.  

Substituting Eq.~(\ref{kinetic_sol}) into Eq.~(\ref{current}), for the quasiparticle contribution to the current we finally obtain
\begin{eqnarray}
j_n = \frac{e^2 N_F v_F}{\pi} \Bigl(V -  \frac{\dot h_y d}{e v_F}\Bigr).
\label{current_normal_2}
\end{eqnarray}

It is seen that in the presence of magnetization dynamics there is  an electromotive force ${\cal E} = \dot h_y d /(e v_F)$ in the TI resulting from the emergent electric field induced due to the simultaneous presence of the time-dependent exchange field and spin-momentum locking. 

\section{Electrical response to magnetization dynamics}

\subsection{Open circuit}

Now let us assume that the domain structure of the strip is moved along $y$-axis. Let us consider the setup presented in Fig.~\ref{sketch}(a) with an insulating ferromagnet and assume that an electric current is applied between the S(N) leads. The total electric  current $j$ through the junction is a sum of the supercurrent contribution Eq.~(\ref{Josephson_CPR}) and the normal current Eq.~(\ref{current_normal_2}) flowing via the TI surface states. It can be rewritten as follows:
\begin{eqnarray}
j=j_c \sin (\chi - \chi_0)+\frac{1}{2eR_N}(\dot \chi - \dot \chi_0)
\label{current_total}
\end{eqnarray}
For the open circuit $j=0$ and the solution of Eq.~(\ref{current_total}) is $\chi (t) = \chi_0(t)$. Therefore, the voltage generated at the junction due to magnetization dynamics in the ferromagnet is
\begin{eqnarray}
V_x = \frac{\dot \chi}{2e} = \dot h_y d/e v_F,
\label{voltage}
\end{eqnarray}
where we denote the voltage at the junction presented in Fig.~\ref{sketch}(a) as $V_x$. The same physical picture is valid for the junction presented in Fig.~\ref{sketch}(b), but the voltage $V_y$ at this junction is determined as $V_y = \dot h_x d/e v_F$. In the both cases the voltage is determined by the dynamics of the magnetization component perpendicular to the current direction. It is the same as for superconducting, so as for nonsuperconducting leads at $j = 0$ and is determined only by the emf. This is the consequence of the fact that the emf and the anomalous phase shift are manifestations of the same gauge vector potential, which is determined only by the spin-momentum locking and magnetization and is not influenced by superconductivity. 

The effect of this voltage generation can be used for electrical detection of magnetization dynamics. Measuring the voltages at the two junctions (sketched in panels (a) and (b) of Fig.~\ref{sketch}), which are called by "detectors" further and attached to one and the same ferromagnetic strip, one can obtain the full time dependence of the in-plane magnetization components $M_x(t)$ and $M_y(t)$ in the ferromagnet. The particular example of the corresponding numerical simulation is demonstrated below.

To be concrete we consider the magnetic field-driven DW motion along the ferromagnet strip. The field-induced and current induced DW motions have been widely investigated both theoretically and experimentally \cite{Slonczewski1972,Schryer1974,Ono1999,Atkinson2003,Saitoh2004,Nakatani2001,Thiaville2002,Nakatani2003,Beach2005,Fukumoto2005,Hayashi2006,Hayashi2007,Beach2006,Klaui2005,Grollier2003,Yamaguchi2004,Thomas2006,Vernier2004,Tsymbal2016}.  In principle, the electric current $j \neq 0$ flowing via the detector arranged as shown in Fig.~\ref{sketch}(b) can generate an additional torque on the magnetization\cite{Yokoyama2010,Yokoyama2011,Mahfouzi2012,Chen2014}. For the open circuit considered here this contribution is absent, and we also neglect it in the next subsection, where we assume $j \neq 0$, because we assume the current via the detector to be small and neglect its additional minor contribution to the external torque moving the DW.  We find $\bm M(y,t)$ numerically from the Landau-Lifshitz-Gilbert (LLG) equation 
\begin{eqnarray}
\frac{\partial\bm M}{\partial t} = -\gamma \bm M \times H_{eff} + \frac{\alpha}{M_s} \bm M \times \frac{\partial\bm M}{\partial t},
\label{LLG}
\end{eqnarray}
where $M_s$ is the saturation magnetization, $\gamma$ is the gyromagnetic ratio and $H_{eff}$ is the local effective field
\begin{eqnarray}
\bm H_{eff} = \frac{H_K M_y}{M_s}\bm e_x + \frac{2A}{M_s^2}\nabla_y^2 \bm M - 4 \pi M_z \bm e_z + H_{ext} \bm e_x~~~~~~
\label{H_eff}
\end{eqnarray}
$H_k$ is the anisotropy field, along the $x$-axis, $A$ is the exchange constant, $\bm H_{ext}$ is the external field and the self-demagnetization field $4\pi M_z$ is included. For numerical calculations we use the material parameters of YIG thin films \cite{Mendil2019}: $H_K \sim 0.5 ~Oe$, $4 \pi M_s = 1000~ Oe$, $\alpha \sim 0.01$ and the domain wall width $d_W = 1\mu m$.

\begin{figure}[htb!]
\begin{minipage}[b]{\linewidth}
 \centerline{\includegraphics[clip=true,width=2.8in]{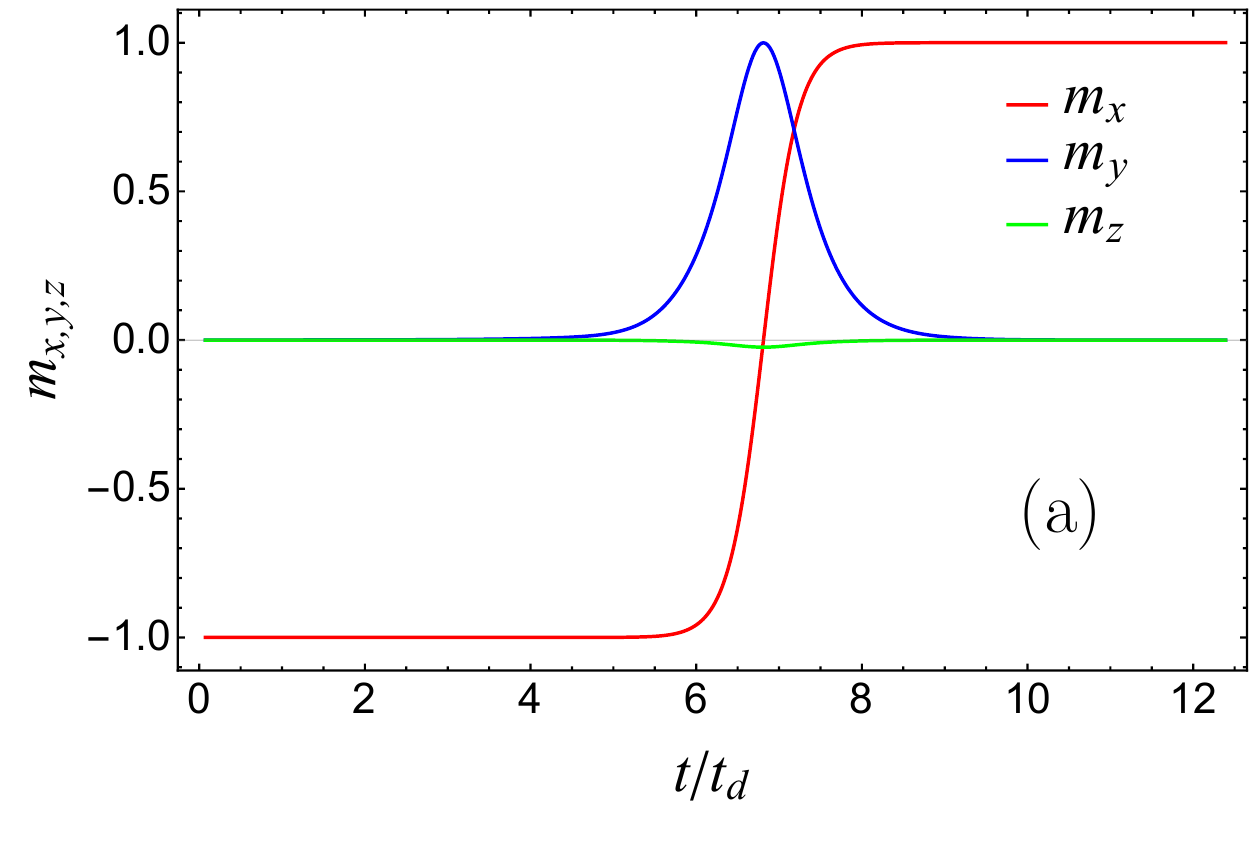}}
 \end{minipage}\hfill
\begin{minipage}[b]{\linewidth}
 \centerline{\includegraphics[clip=true,width=2.8in]{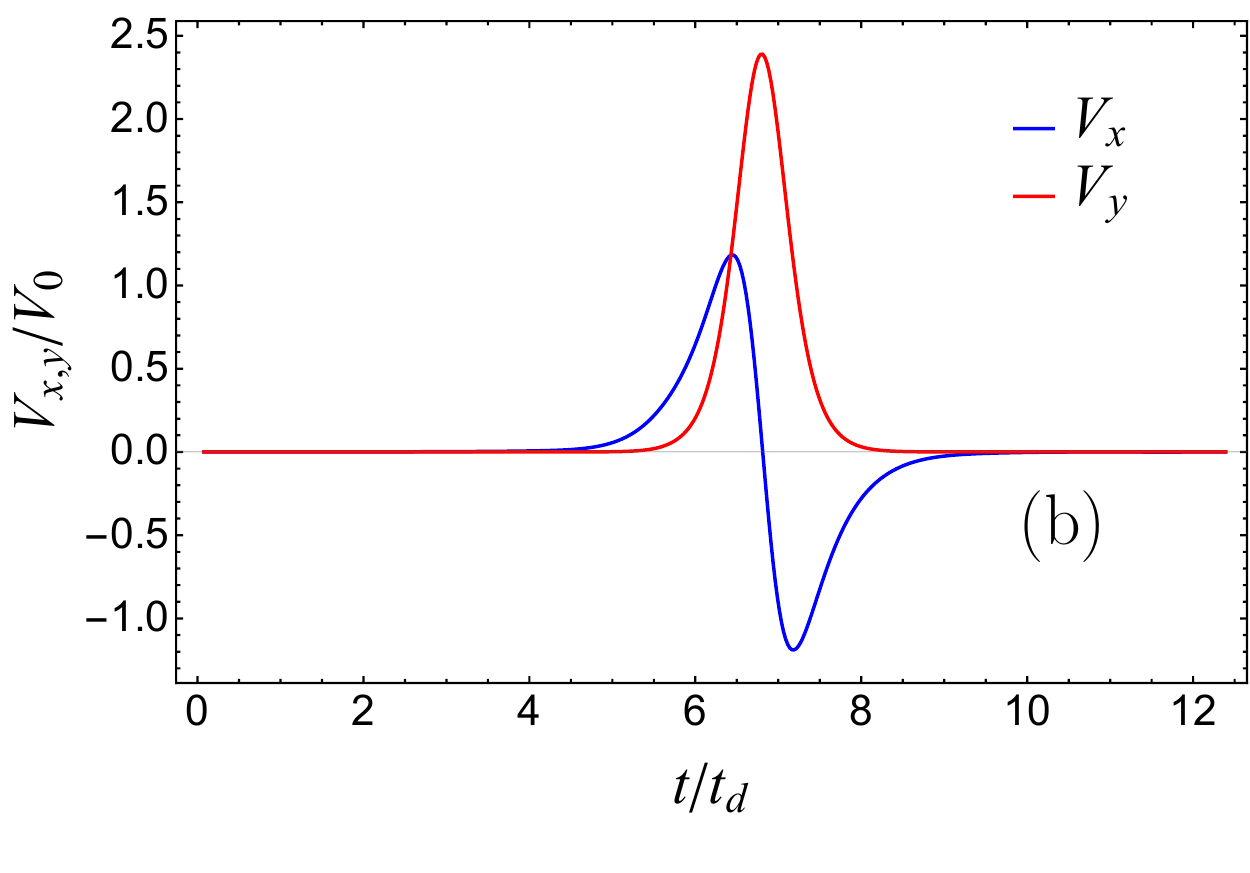}}
 \end{minipage}\hfill 
 \caption{(a) Magnetization components $m_{x,y,z}=M_{x,y,z}/M_s$ in  the region of the detectors and (b) voltages $V_{x,y}$ as functions of time. $H_{ext}=0.003M_s$, for other parameters of the numerical calculation see text. 
        }
       \label{magnetization1}
 \end{figure}

\begin{figure}[htb!]
\begin{minipage}[b]{\linewidth}
 \centerline{\includegraphics[clip=true,width=2.8in]{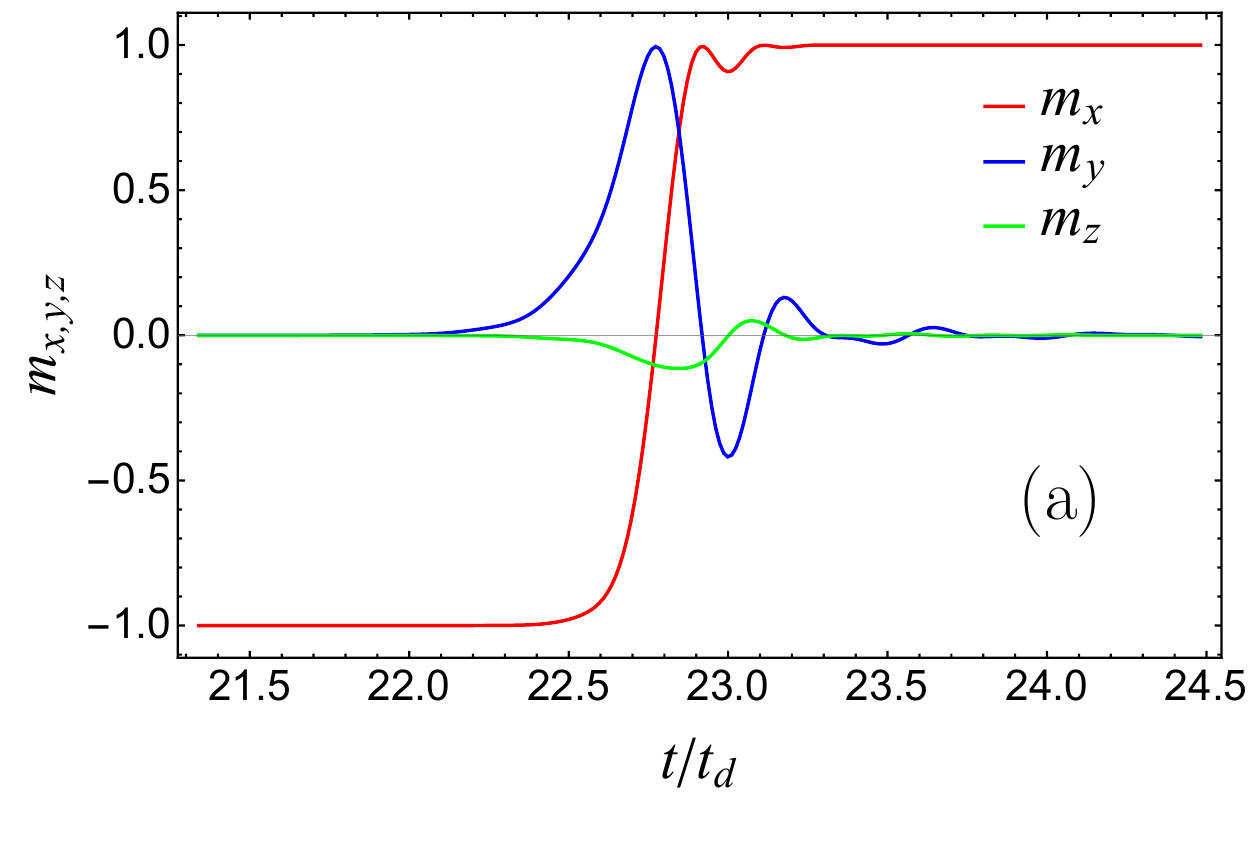}}
 \end{minipage}\hfill
\begin{minipage}[b]{\linewidth}
 \centerline{\includegraphics[clip=true,width=2.8in]{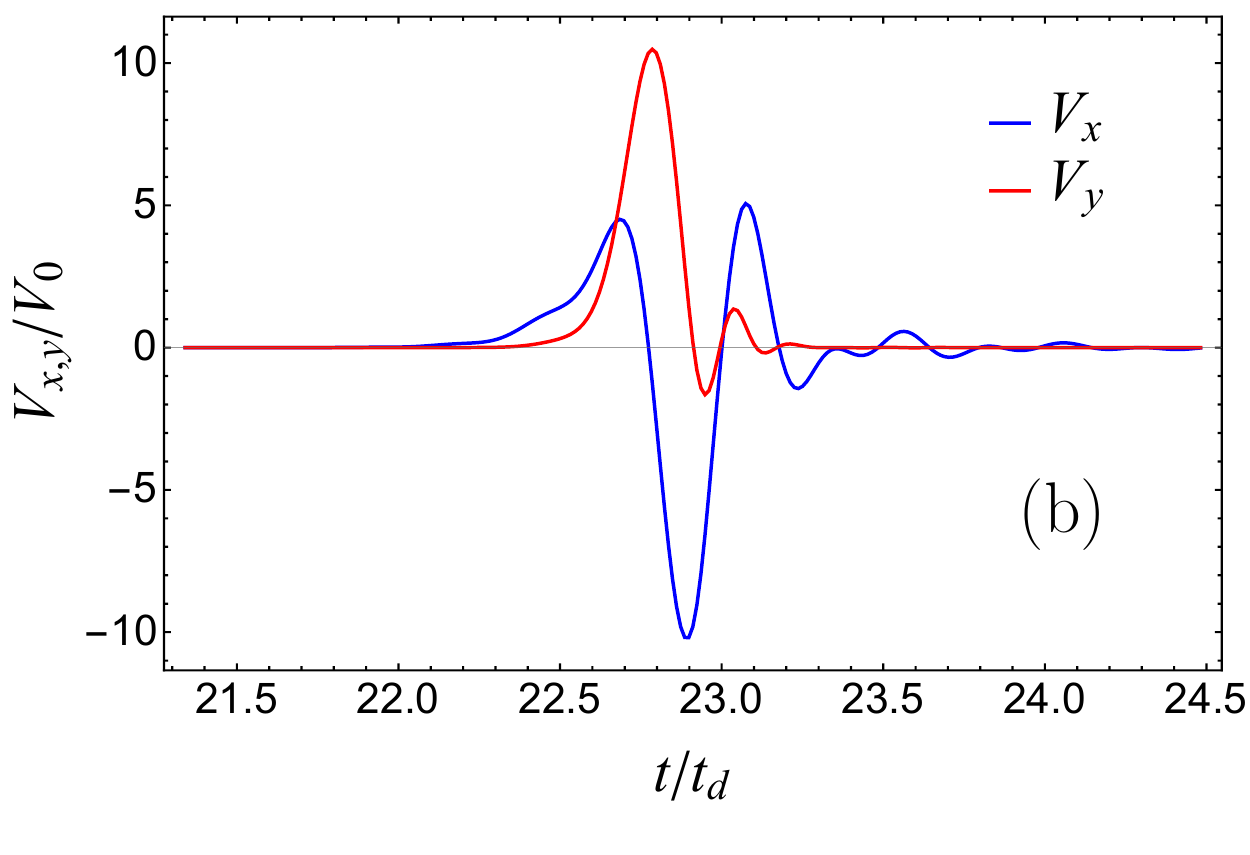}}
 \end{minipage}\hfill 
 \caption{(a) Magnetization components $m_{x,y,z}=M_{x,y,z}/M_s$ at the detectors and (b) voltages $V_{x,y}$ as functions of time. $H_{ext}=0.02M_s$, the detectors are located in the spatial region, where the DW motion is steady. 
        }
       \label{magnetization2}
 \end{figure}

As it has already been defined above the voltage, measured by the detector shown in Fig.~\ref{sketch}a(b) is denoted by $V_{x(y)}$ and probes $\dot M_{y(x)}$. The result of the numerical calculation of $M_{x,y,z}(t)$ at the detectors is presented in panels (a) of Figs.~\ref{magnetization1}-\ref{magnetization2}, while the corresponding dependencies $V_{x,y}(t)$ are presented at panels (b) of these figures. As the characteristic size of the region containing the both detectors $\sim 100nm$ is assumed to be much smaller than the characteristic scale of magnetic inhomogeneity $d_W$, we can safely use our Eq.~(\ref{voltage}), which is strictly valid for a homogeneous magnetization, for calculation of the voltage at the detector. 

Fig.~\ref{magnetization1} demonstrates the results for small enough external applied field $H_{ext}=0.003M_s$. This field is below the Walker's breakdown field \cite{Schryer1974} and in this regime the DW moves keeping its initial plane structure. Having at hand $V_{x,y}(t)$ it is possible to restore the time-dependent structure of the moving wall at the detector point according to the relation
\begin{equation}
\Delta M_{x,y}(t)/{M_s} = (1/V_0t_d) \int \limits_{t_i}^t V_{y,x}dt,
\label{magnetization_restore}
\end{equation}
where $V_0 = d h_{eff}/(|e|v_Ft_d)$ is the natural unit of the voltage induced  at a given detector. The characteristic time of magnetization variation can be obtained from the LLG equation and takes the form $t_d = 1/(4 \pi \alpha \gamma M_s)$. Taking the material parameters of YIG thin films we obtain that $t_d \sim 0.5 \times 10^{-8}s$. 

Fig.~\ref{magnetization2} represents the results for higher applied field $H_{ext} = 0.02 M_s$, which exceeds the Walker's breakdown field. In this regime the DW initial plane shape is not preserved during the motion, what is reflected in the oscillations of all the magnetization components, as it is seen in Fig.~\ref{magnetization2}(a). In this regime the magnetization at the detector point as a function of time can also be found as described above.

Observation of a single DW traveling  through the detector region gives a way to find experimentally the effective exchange field $h_{eff}$ induced by the ferromagnet in the surface layer of the TI. Indeed, in this case $|\Delta M_x/M_s| = 2$ after the DW passes through the detector region. Then $h_{eff}$ can be found from Eq.~(\ref{magnetization_restore}) as follows:
\begin{eqnarray}
h_{eff} = |(e v_F/2 d) \int \limits_{t_i}^{t_f} V_{y}dt|,
\label{h_find}
\end{eqnarray}
where $V_y$ should be integrated over the whole time region, when the voltage is nonzero.

Typical values of the voltage induced at the detectors are of the order of $V_0$. It is difficult to give an accurate a-priori estimate of $V_0$ because there are no reliable experimental data for $h_{eff}$. However, basing on the experimental data discussed above and assuming  $h_{eff} \sim 20-1000 K$ and $d/(v_F t_d) \sim 10^{-4}$, we obtain $V_0 \sim 0.2-10 \mu V$.

If the ferromagnet is metallic, there is also an additional normal current flowing via the ferromagnet. The resistance of the ferromagnet is typically much smaller than the resistance of the TI surface states. For this reason the voltage induced at the junction due to the presence of the emf in the TI is suppressed by the factor $R_F/(R_F+R_N)$, where $R_{F(N)}$ is the resistance of the ferromagnet (TI surface states). Therefore, metallic ferromagnets are not good candidates for measuring the discussed effect. In addition, as it was already mentioned, the time-dependent spin texture of a ferromagnet also gives rise to emergent spin-dependent electric and magnetic fields and, consequently, to an additional parasitic voltage in it. This voltage can interfere with the emf generated in the TI and the resulting effect is quite complicated.

In principle, a similar structure of the emf can also be obtained for systems where a topological insulator is replaced by a material with Rashba spin-orbit coupling or if the Rashba spin-orbit coupling is an internal property of the ferromagnetic film arising, for example, from  structural asymmetry in the $z$-direction. However, we consider the TI-based systems to be a more preferable variant because here the effect should be stronger. The emf generated in $x$-direction is also predicted to be proportional to $\dot h_y$ in Rashba spin-orbit based junctions\cite{Kim2012,Tatara2013,Yamane2013,Rabinovich2019}, but in addition it should contain a reducing factor $\Delta_{so}/\varepsilon_F \sim \alpha_R/\hbar v_F$. This factor can be estimated by taking a realistic value of $\alpha_R \sim 10^{-10} eV m$ \cite{Henk2004,Ast2007} for the interfaces of heavy-metal systems. Then $\Delta_{so}/\varepsilon_F \sim 0.1$ if one assumes $v_F \sim 10^6 ms^{-1}$. Therefore, it reduces the value of the emf, which is important for the magnetization detection.

\subsection{Response in the presence of the applied electric current}

Now we turn to the case when the constant electric current $j$ is applied to the junction. In this case the voltage $V=\dot \chi/2e$ should be found from Eq.~(\ref{current_total}). If $j_c$ does not depend on time, the solution of this equation represents the well-known voltage $V_J(t)$ of the ac Josephson effect \cite{Schmidt}  shifted due to the presence of magnetization dynamics $V = V_J(t)+\dot \chi_0/2e$. This leads to the appearance of the nonzero resistance of the $IV$-characteristics at $j<j_c$ and an additional resistance at $j>j_c$ \cite{Rabinovich2019}. However, the specific feature of the TI-based system is the strong dependence of $j_c$ on the effective exchange field component parallel to the current direction, what results in the strong dependence of $j_c$ on time in the presence of magnetization dynamics.

\begin{figure}[htb!]
\begin{minipage}[b]{\linewidth}
 \centerline{\includegraphics[clip=true,width=2.8in]{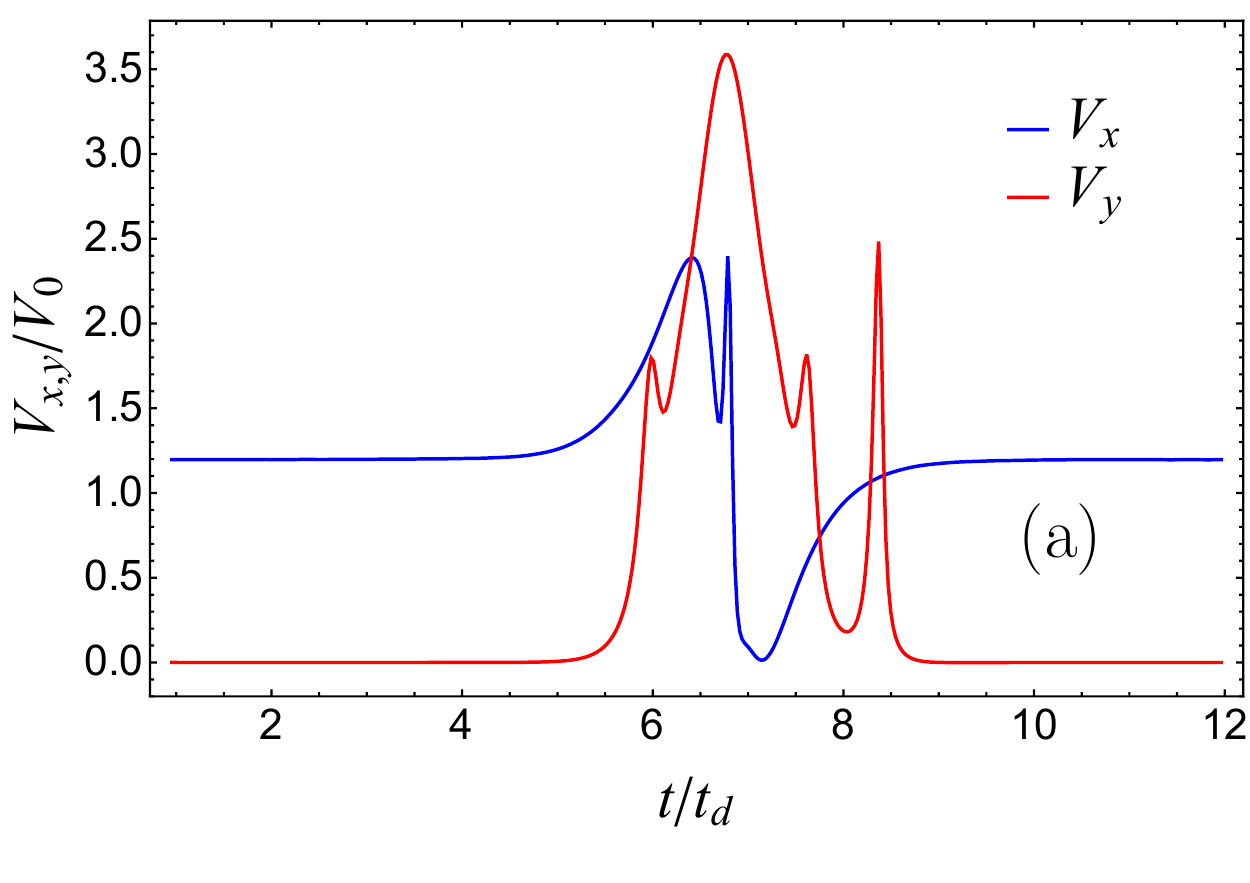}}
 \end{minipage}\hfill
\begin{minipage}[b]{\linewidth}
 \centerline{\includegraphics[clip=true,width=2.8in]{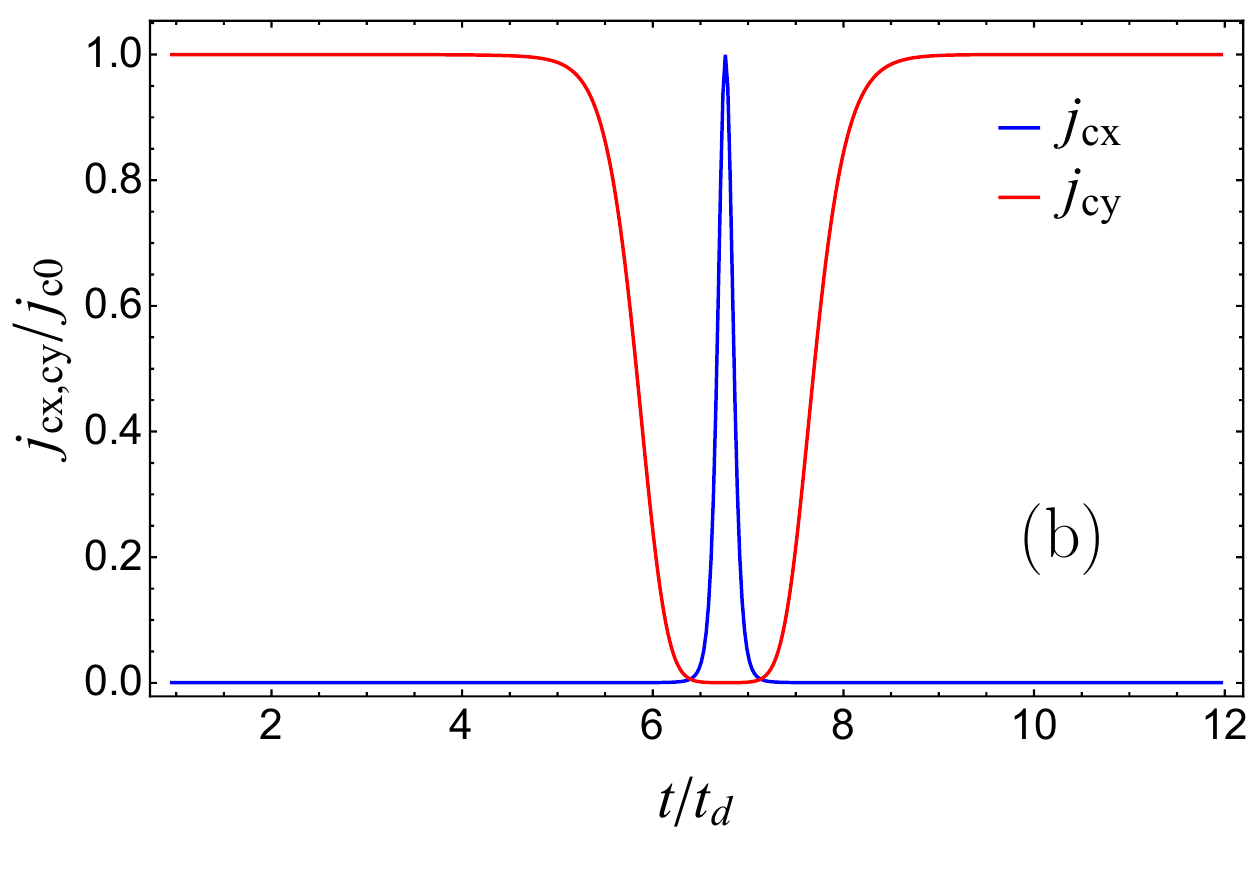}}
 \end{minipage}\hfill 
 \caption{(a) Voltages $V_{x,y}$ as functions of time. (b) Critical currents $j_{cx}$ and $j_{cy}$ as functions of time. $j=0.9j_{c0}$, $h_{eff}=100K$, $j_{c0}R_N = 1\mu V$, $V_0 = 0.75 \mu V$. The other parameters are the same as for Fig.~\ref{magnetization1}. 
        }
       \label{magnetization1_jne0}
 \end{figure}

\begin{figure}[htb!]
\begin{minipage}[b]{\linewidth}
 \centerline{\includegraphics[clip=true,width=2.8in]{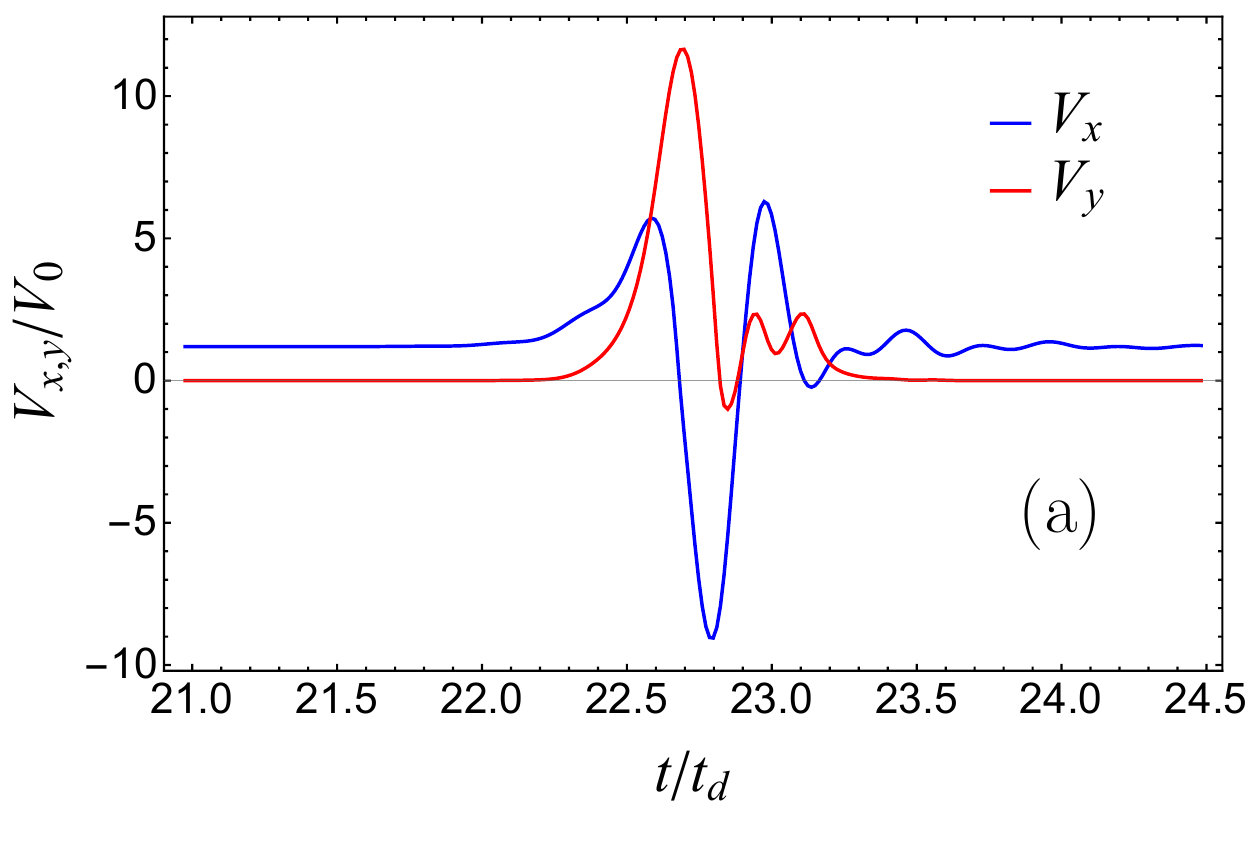}}
 \end{minipage}\hfill
\begin{minipage}[b]{\linewidth}
 \centerline{\includegraphics[clip=true,width=2.8in]{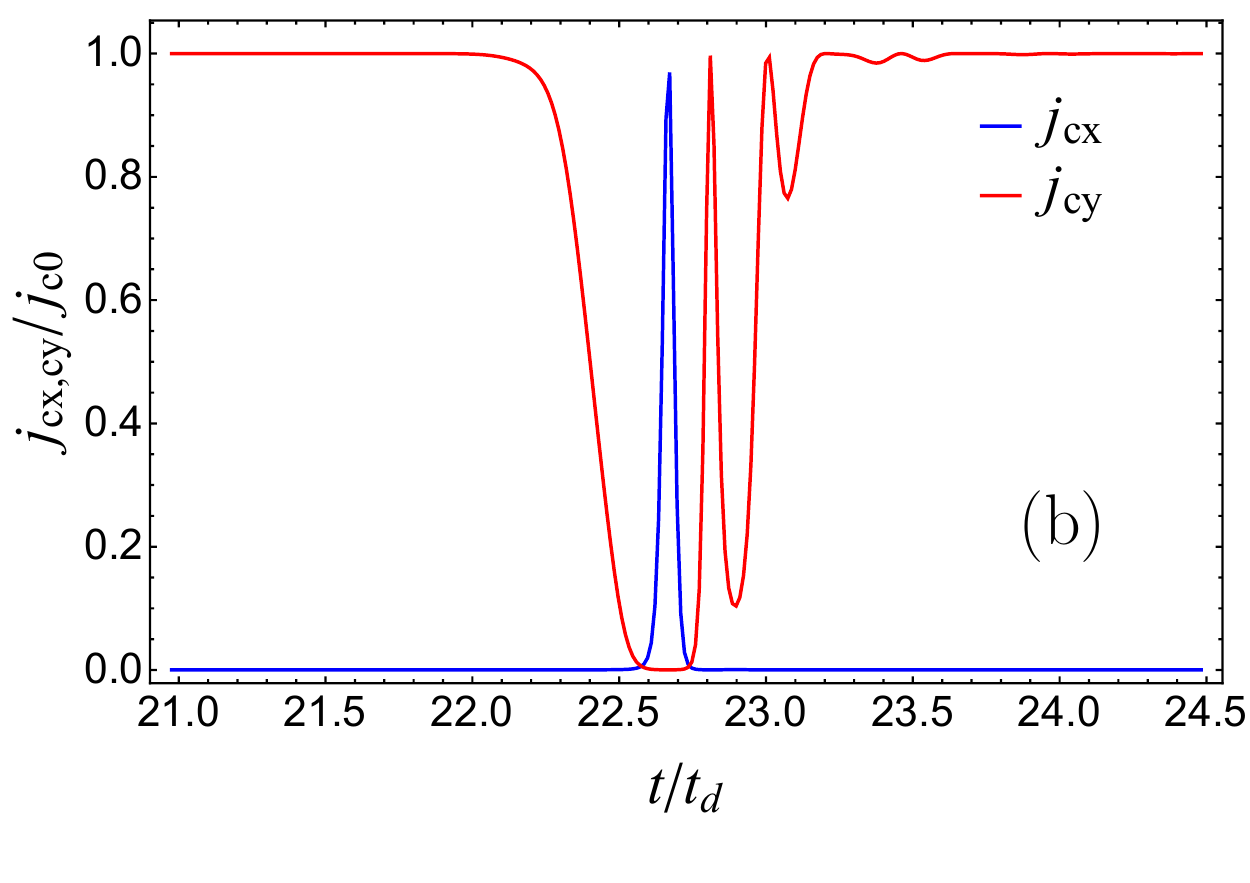}}
 \end{minipage}\hfill 
 \caption{(a) Voltages $V_{x,y}$ as functions of time. (b) Critical currents $j_{cx}$ and $j_{cy}$ as functions of time. $j=0.9j_{c0}$, $h_{eff}=100K$, $j_{c0}R_N = 1\mu V$, $V_0 = 0.75 \mu V$. The other parameters are the same as for Fig.~\ref{magnetization2}. 
        }
       \label{magnetization2_jne0}
 \end{figure}

In Figs.~\ref{magnetization1_jne0}-\ref{magnetization2_jne0} we present the results of the electrical response of the Josephson detectors on the moving DW  under the constant applied current flowing via the detectors. These figures are plotted for $h_{eff} = 100 K$, when the dependence of the critical current on magnetization orientation is given by the blue solid curve in Fig.~\ref{current_orientation}. We take the value of the applied current $j = 0.9j_{c0}$. Depending on the particular magnetization orientation this value of the applied current can be lower as well as higher than the critical current of the junction for a given orientation, see Fig.~\ref{current_orientation}. Figs.~\ref{magnetization1_jne0}a-\ref{magnetization2_jne0}a demonstrate the voltages induced at the detectors as functions of time, while the critical currents of the detectors are shown in panels (b). We denote the critical current of the junction shown in Fig.~\ref{sketch}a(b) as $j_{cx(cy)}$. Fig.~\ref{magnetization1_jne0} corresponds to the parameters of Fig.~\ref{magnetization1}a when the magnetic field driven the DW motion is below the Walker's breakdown, and Fig.~\ref{magnetization2_jne0} represents the case when it exceeds the Walker's breakdown field and the magnetization of the moving DW oscillates, as in Fig.~\ref{magnetization2}a. 

According to Eq.~(\ref{critical_current_T}) the critical current $j_{cx(cy)}$ depends on time via the dependence of $h_{x(y)}$. This dependence is clearly seen in Figs.~\ref{magnetization1_jne0}b-\ref{magnetization2_jne0}b. When there is no DW inside the detector region, $j_{cy}$ is maximal, because there is only $h_{eff,x}\neq 0$, that is the component of the effective exchange field perpendicular to the Josephson current direction, which does not suppress the value of the critical current. At the same time $j_{cx}$ is fully suppressed by this effective field component, because for this detector it is parallel to the current. When a DW passes through the detector region the situation changes: $j_{cy}$ is suppressed by nonzero $h_{eff,y}$, and $j_{cx}$ is restored due to the decrease of $h_{eff,x}$ in the region occupied by the DW.

The $j_{cy(cx)}(t)$ dependence on time manifests itself in the voltage induced at the corresponding Josephson junction. As it is seen from Figs.~\ref{magnetization1_jne0}a-\ref{magnetization2_jne0}a, voltage $V_x$ is nonzero when there are no DWs in the detector region. This detector is in the resistive state because its critical current is less then the externally applied current. When a DW travels via the detectors region we observe as $V_{x}$, so as $V_y$ voltage pulses. These pulses are the results of two different effects: (i) the contribution due to emf, the same as in the open circuit discussed above and (ii) the contribution due to the dependence of the critical current on time, which leads to the time dependence of the phase difference $\chi$ between the superconductors. For the example under consideration $j_{cy}$ becomes lower than the applied current when the DW passes through the detector region. This leads to the appearance of the Josephson oscillations, which are seen in Fig.~\ref{magnetization1_jne0}a. The picture of the oscillating DW moving under a field higher than the Walker's breakdown, presented in Fig.~\ref{magnetization2_jne0}, is more complicated. The magnetization oscillations are clearly seen in the dependence of the critical current on time. These oscillations manifest themselves in the dependencies $V_{x,y}$ and for the chosen parameters  interfere with Josephson oscillations.

 \begin{figure}[htb!]
\begin{minipage}[b]{\linewidth}
 \centerline{\includegraphics[clip=true,width=2.8in]{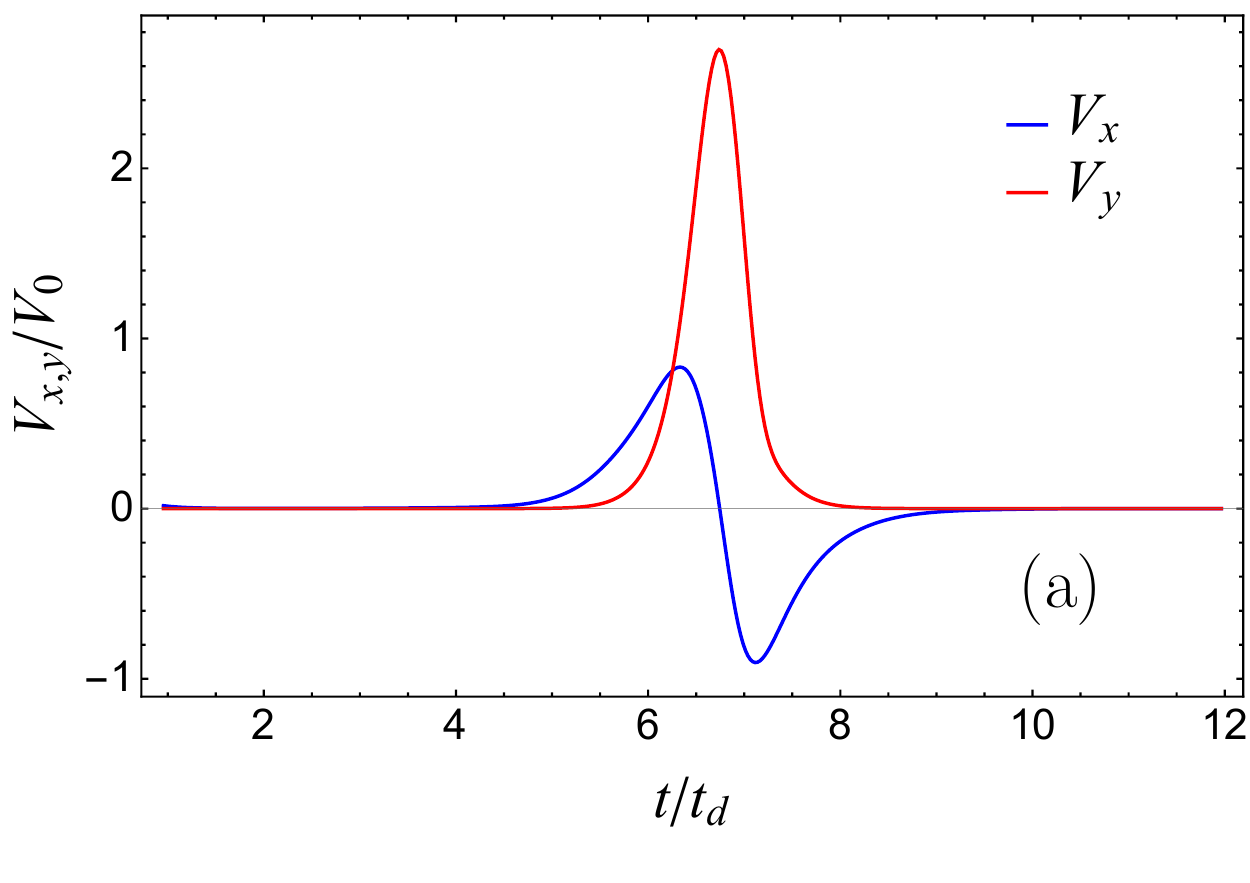}}
 \end{minipage}\hfill
\begin{minipage}[b]{\linewidth}
 \centerline{\includegraphics[clip=true,width=2.8in]{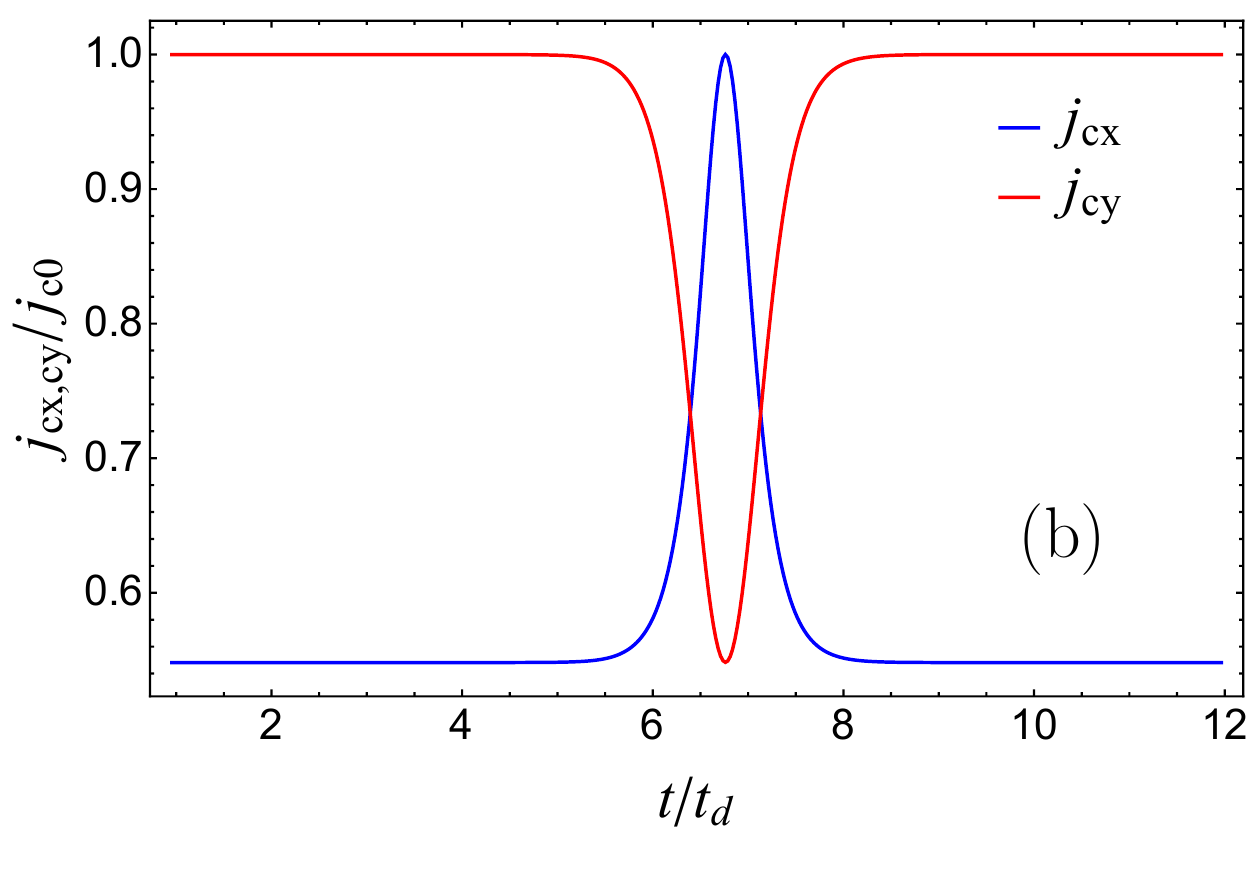}}
 \end{minipage}\hfill 
 \caption{(a) Voltages $V_{x,y}$ as functions of time. (b) Critical currents $j_{cx}$ and $j_{cy}$ as functions of time. $j=0.5j_{c0}$, $h_{eff}=20K$, $j_{c0}R_N = 1\mu V$, $V_0 = 0.15 \mu V$. The other parameters are the same as for Fig.~\ref{magnetization1}. 
        }
       \label{magnetization1_j}
 \end{figure} 

\begin{figure}[htb!]
\begin{minipage}[b]{\linewidth}
 \centerline{\includegraphics[clip=true,width=2.8in]{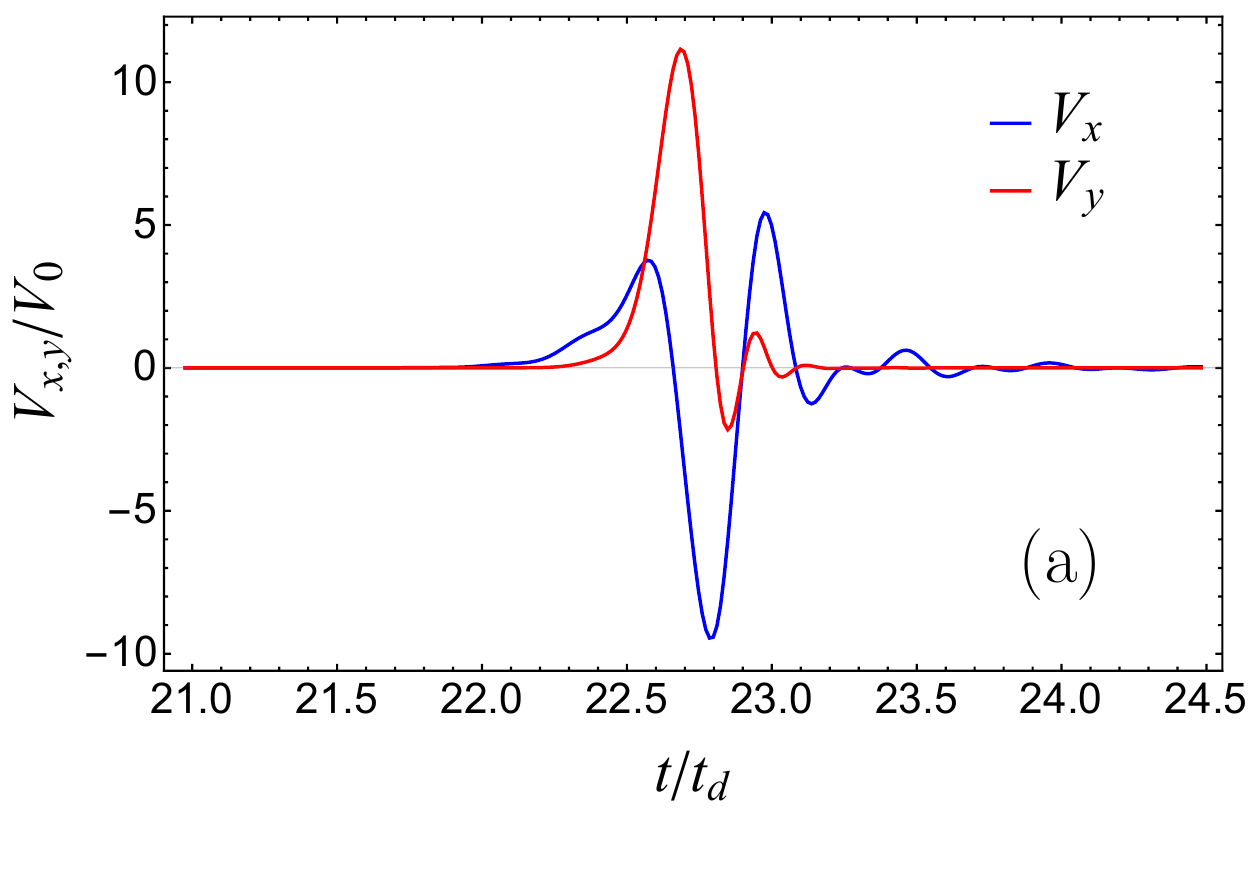}}
 \end{minipage}\hfill
\begin{minipage}[b]{\linewidth}
 \centerline{\includegraphics[clip=true,width=2.8in]{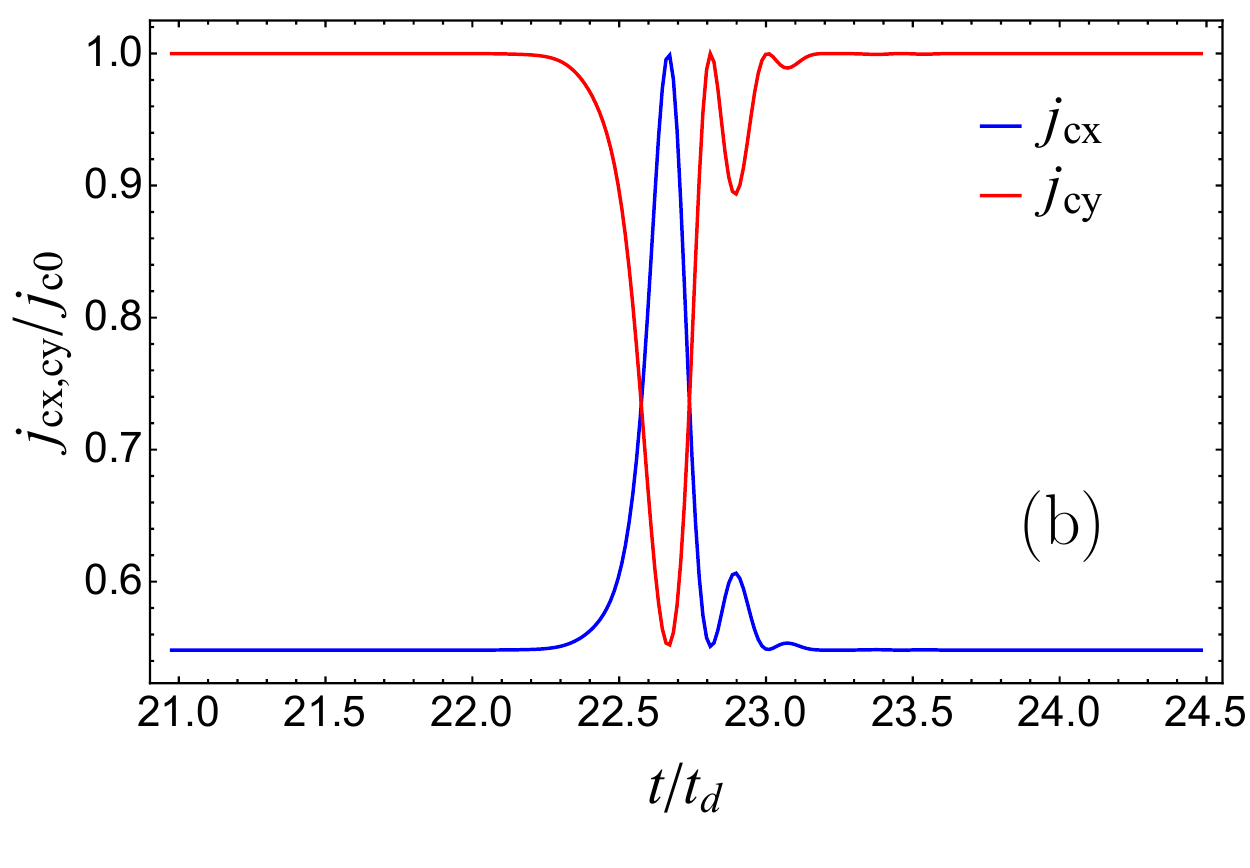}}
 \end{minipage}\hfill 
 \caption{(a) Voltages $V_{x,y}$ as functions of time. (b) Critical currents $j_{cx}$ and $j_{cy}$ as functions of time. $j=0.5j_{c0}$, $h_{eff}=20K$, $j_{c0}R_N = 1\mu V$, $V_0 = 0.15 \mu V$. The other parameters are the same as for Fig.~\ref{magnetization2}. 
        }
       \label{magnetization2_j}
 \end{figure} 
 
In Figs.~\ref{magnetization1_j}-\ref{magnetization2_j} we consider another regime, when the applied current is less than the critical current of the detectors for any magnetization orientation. The parameters correspond to the case of lower effective exchange field $h_{eff} = 20K$, corresponding to the solid red curve in Fig.~\ref{current_orientation}. It is seen that in the absence of the moving DW inside the detector region the both detector junctions are in the dissipationless regime. The voltage pulses occur in $V_{x,y}$ when the DW passes through the detectors. In this case there are no Josephson oscillations in $V_{x,y}$, as it is seen in Fig.~\ref{magnetization1_j}a. The shape of the voltage pulses is very close to the shape obtained for the open circuit, however the amplitudes of the pulses are different due to the additional contribution from the dependence of the critical current on time. Fig.~\ref{magnetization2_j} demonstrates the electrical response of the detectors to the oscillating motion of the DW above the Walker's breakdown. These oscillations again manifest themselves in the dependence of the critical currents $j_{cx(cy)}$ on time. In the present case they directly appear in the oscillating behavior of $V_{x,y}$ without a contamination by the Josephson oscillations because the applied current does not exceed the critical current of the detectors. 

As we can see, the experiment under the applied current is less preferable for detection of the magnetization dynamics because one cannot extract a pure emf signal due to the time dependence of the critical current. However, it might be of interest for experimental determination of $j_c(t)$. Indeed, simultaneous measurements of the electrical voltage at the junction in the open circuit and under the applied current allows for the calculation of  $j_{c}(t)$ and, consequently, $j_c(M_x, M_y)$ according to Eq.~(\ref{current_total}).

In principle, the contributions to the voltage from the emf and from the time-dependence of the critical current can be separated. For example, if one considers the dynamics of a DW wall with perpendicular anisotropy \cite{Heinrich1993}, located in the $(y,z)$-plane, then for the detector presented in Fig.~\ref{sketch}b there is no in-plane exchange field component, which is perpendicular to the current. Therefore, the emf contribution does not occur at this detector. At the same time, the component of $h_{eff}$ parallel to the current is absent at the detector in Fig.~\ref{sketch}a. Consequently, the critical current does not depend on time for this detector, and the emf is the only contribution to the voltage. 
 
\section{Conclusions}

The electrical response of the S/3D TI-F/S Josephson junction to magnetization dynamics was studied and compared to the electrical response of the junction with nonsuperconducting leads. In 3D TI/F hybrid structures spin-momentum  locking of the 3D TI conducting surface states in combination with the induced  magnetization leads to the appearance of a gauge  vector  potential. In the presence of magnetization dynamics the gauge vector potential becomes time-dependent and generates an electromotive force. In both cases of superconducting and nonsuperconducting leads this emf generates a voltage between the leads. For an open circuit this voltage  is  the  same  for both normal  and  superconducting  leads  and  allows  for  electrical  detection  of magnetization dynamics, a structure of a time-dependent magnetization and a measurement of the effective exchange field. In the presence of the applied current the electrical response of the Josephson junction contains additional contribution from the time dependence of the critical Josephson current, which comes from the strong dependence of the critical current on the magnetization orientation. In particular geometries it complicates quantitative detection of the exact shape of the time-dependent magnetization texture. At the same time, it can be used for experimental investigation of the dependence of the critical Josephson current via the 3D TI proximitized by a ferromagnetic insulator on the exchange field orientation.

\section{Acknowledgments}

The authors are grateful to M.A. Silaev for discussions. The work was supported by the Russian Science Foundation Project No. 18-72-10135.

\end{document}